# When Two Worlds Collide: Using Particle Physics Tools to Visualize the Limit Order Book

Marjolein E. Verhulst[1,2], Philippe Debie[1,2], Stephan Hageboeck[3], Joost M.E. Pennings[1,4,5], Cornelis Gardebroek[1], Axel Naumann[3], Paul van Leeuwen[2], Andres A. Trujillo-Barrera[6], Lorenzo Moneta[3]

[1] Wageningen University & Research, Wageningen, the Netherlands

[2] Wageningen Economic Research, Den Haag, the Netherlands

[3] European Organization for Nuclear Research (CERN), Meyrin, Switzerland

[4] Maastricht University, Maastricht, the Netherlands

[5] University of Illinois at Urbana-Champaign, Urbana, IL, United States of America

[6] University of Idaho, Moscow, ID, United States of America

**Correspondence**

Marjolein Verhulst, Marketing and Consumer Behaviour Group, Wageningen University & Research, Hollandseweg 1, 6706 KN, Wageningen, the Netherlands. Email: marjolein.verhulst@wur.nl

Funding information

This research did not receive any specific grant from funding agencies in the public, commercial, or not-for-profit sectors.

Declarations of interest

None.

Data Availability Statement

The data that support the findings of this study are available from the Chicago Mercantile Exchange. Restrictions apply to the availability of these data, which were used under license for this study. The data are available from the authors with the permission of the Chicago Mercantile Exchange.

Abstract

We introduce a methodology to visualize the limit order book (LOB) using a particle physics lens. Open-source data-analysis tool ROOT, developed by CERN, is used to reconstruct and visualize futures markets. Message-based data is used, rather than snapshots, as it offers numerous visualization advantages. The visualization method can include multiple variables and markets simultaneously and is not necessarily time dependent. Stakeholders can use it to visualize high-velocity data to gain a better understanding of markets or effectively monitor markets. In addition, the method is easily adjustable to user specifications to examine various LOB research topics, thereby complementing existing methods.

Acknowledgements

The authors would like to thank an anonymous referee and the Editor, Dr. Webb, for their constructive comments and suggestions. We thank the European Organization for Nuclear Research (CERN, Geneva) for extending their analytical ROOT system and providing data storage and computing power for this research. We are grateful to the Chicago Mercantile Exchange Foundation for providing market depth data for all their futures and options for the full year of 2015, containing all market messages, timestamps and levels of the limit orderbook. We would like to thank the Office for Futures and Options Research at the University of Illinois at Urbana-Champaign, the Commodity Risk Management Expertise Centre (CORMEC), the Dutch Authority for the Financial Markets (AFM), Euronext and the Dutch National Bank (DNB) for their constructive feedback and insights on previous versions of this manuscript. Lastly, we would like to thank SURF SARA for granting access to the national computing and data storage environment (Grud HPC Cloud Beehub, Hadoop).

1. **Introduction**

The transition of financial and commodity exchanges from physical trading venues hosting open outcry auctions to predominantly electronic trading platforms (Hirsch, Cook, Lajbcygier, & Hyndman, 2019), precipitated two major changes to financial markets. First, the limit order books (LOB) of electronic trading platforms became partially visible to market participants. Second, the shift towards electronic trading platforms has transformed trading, in that algorithms can now trade, which has led to the emergence of high-frequency traders (HFT). Indeed, HFT and other forms of algorithmic trading now account for the majority of market turnover (Hirsch et al., 2019) and has introduced new challenges such as "Flash Crashes" (Aldridge & Krawciw, 2017; Bayraktar & Munk, 2018; CFTC-SEC, 2010; Golub, Keane, & Poon, 2012; Kirilenko, Kyle, Samadi, & Tuzun, 2017; Menkveld & Yueshen, 2019).

These changes create new challenges and opportunities for academics, regulators and industry participants alike, since they are faced with new high-frequency data that is more detailed and richer than ever before[1]. One approach to adjusting to the changes is to visualize the data. Visualization of the LOB helps stakeholders to provide the context of the market in which traders perform their actions and, hence, improve the understanding of the market they are analyzing and/or operating in and allows them, among others, to detect and identify anomalies. However, since LOB data is voluminous, complex in terms of structure and arriving at high frequencies, there is a need for a new way of thinking about storing, processing, visualizing and analyzing such data. This study attempts to address this issue by advocating the application of a visualization methodology commonly used in the particle physics literature to finance (see Appendix A). Specifically, we try to answer the research question: to what extent can particle physics methodologies be used to visualize LOB data?

---

[1] For example, it takes the Securities and Exchange Commission five months to process and analyze two hours of LOB data (Gai et al., 2014).

The objective of this paper is twofold: first, to offer a novel visualization of the LOB, that is customizable by user specifications, using particle physics visualization tools. Second, to illustrate how the proposed visualization tool is easily adjustable to study various topics in LOB research, e.g. liquidity. Specifically, we use the open source data-analysis tool for high-energy physics ROOT, developed by CERN, among others, to reconstruct and visualize LOBs (Brun & Rademakers, 1997; CERN, 2018b).

This paper contributes to the existing literature by providing a visualization methodology that complements the existing visualizations with the following novel features: 1) LOB data can be visualized in different ways; either with time or messages on the x-axis, i.e. using "snapshots" or market activity; 2) the ability to visualize an extensive number of variables simultaneously; 3) the visualization of more complex concepts with separate but connected variables, such as liquidity; and 4) the visualization of multiple markets simultaneously. The methodology provides the ability to render LOB data in accordance with user specifications, making it visually easier to comb through LOB data and perform LOB data introspection. In addition, this paper introduces researchers to an open-source toolkit to store, process, generate statistics for, visualize and model LOB data. The focus of this paper is on the processing and visualization features of this toolkit.

## 2. LOB Data and Reconstruction

### 2.1. LOB Data

The LOB helps to explain the behavior of traders and informs theories of market microstructure and behavioral finance (Bhattacharya, Kuo, Lin, & Zhao, 2018; Biais, Bisière, & Spatt, 2010; Brolley, 2020; Buti, Rindi, & Werner, 2017; Chen, Kou, & Wang, 2018; Chordia, Hu, Subrahmanyam, & Tong, 2019; Comerton-Forde, Malinova, & Park, 2018; Dugast, 2018). The shift to electronic trading platforms has fundamentally changed market dynamics. LOB data contains more information and allows, among others, for more comprehensive measurements. For example, previous liquidity measurements took into account only one or two dimensions of liquidity whereas LOB data allows us to measure multiple liquidity dimensions simultaneously (Rösch & Kaserer, 2013). In addition, LOB data is stored according to various protocols, depending on the exchange. The Chicago Mercantile Exchange (CME Group), for example, uses the Financial Information eXchange (FIX) protocol, which provides the messages necessary to construct the LOB (FIXtrading, 2020). This means that academics have to reconstruct the LOB themselves, which may be challenging for various reasons[2]. Although regulators can get a better picture of the market using LOB data, its use also poses major challenges (Paddrik, Haynes, Todd, Scherer, & Beling, 2016): the LOB can be difficult to reconstruct, visualize and analyze as it can be quite complex, due to, for example, various order types and different market microstructures across markets (Paddrik et al., 2016). In addition, high-frequency

---

[2] Academics must understand the FIX protocol and the respective tags and values of messages to be able to reconstruct the LOB. Certain tag-value combinations, for example, act as "if-statements", which require careful consideration when reconstructing the LOB.

trading has a high velocity (nano- or milliseconds) and generates great amounts of data.

## 2.2. LOB Reconstruction: Messages vs. Snapshots

LOB data can be used to address many research questions, as it is usually high-frequency data and, thus, rich in information. Brogaard et al. (2014) and Brogaard et al. (2019) highlight the additional information that is present in the LOB, and provide rationales as to why certain LOB data should be examined more carefully. For example, LOB data provides new information on trading strategies (of both HFTs and non-HFTs); (short-term) volatility; trading behavior around news announcements and imbalances in the LOB (Brogaard et al., 2014, 2019); front-running; manipulative strategies (Brogaard et al., 2019); and price discovery/efficiency in the LOB, including its levels and for different order types (Arzandeh & Frank, 2019; Brogaard et al., 2014, 2019; Cao, Hansch, & Wang, 2009). This paper builds on this notion by examining and comparing information present in LOB messages and in snapshots. Message data contains information the exchange receives about market activity[3]. It does not *contain* the LOB but can be used to *recreate* the LOB (FIXtrading, 2020). To the best of our knowledge, most LOB research does not use message data but already reconstructed LOB data. Studies that do use message data generally fail to provide information about the LOB reconstruction process, with a few exceptions (Arzandeh & Frank, 2019; Erenburg & Lasser, 2009). Moreover, little is known about handling, storing and processing the message data and reconstructing a LOB. The reconstruction process of the LOB is further outlined in Section 3 and Appendix B. This section highlights the differences between using messages and snapshots to re-create a LOB.

---

[3] For example, part of a message can look as follows: 52=20150302150404453 107=ZCK5 269=0 270=392.5 271=8. This message was sent on March 2, 2015 at 15:04:04.453 (52=20150302150404453), it concerns the May 2015 corn futures contract (107=ZCK5) and mentions an update on the bid side of the LOB (269=0), whereby the volume at the price level of 392.5 dollar cents is updated to 8 futures contracts (270=392.5 and 271=8).

Orders arrive irregularly in the LOB, meaning that the messages received by the exchange are irregularly spaced over time. Snapshots of the LOB are needed to arrive at regular time intervals for analysis in a time-series framework. For example, if a snapshot size of one second is used, the last message within the one-second snapshot is used, i.e. only the net effect of all messages within one second is observed. The literature offers no uniform method to achieve the optimal snapshot size. Arzandeh and Frank (2019) calculated the optimal snapshot using the average duration of transaction price changes, following Engle and Russell (1998). Others, however, did not calculate the optimal snapshot size, choosing their intervals rather arbitrarily, anywhere between three seconds (Ito & Yamada, 2018), five seconds (Brogaard & Garriott, 2019), one minute (Gai, Choi, O'Neal, Ye, & Sinkovits, 2014; Hautsch & Horvath, 2019; Sinkovits, Feng, & Ye, 2014; Yao & Ye, 2018), three minutes (Hautsch & Horvath, 2019), five minutes (Kandel, Rindi, & Bosetti, 2012) and thirty minutes (Baruch, Panayides, & Venkataraman, 2017).

The use of snapshots can be problematic for several reasons. First, if the snapshot size is too big, much of the data and information is lost in compression (i.e. into one snapshot), in that the snapshot only shows the aggregation of actions whereas their relative timings are lost. Conversely, if the snapshot size is too small, observations might repeat themselves – e.g. when there is little activity – introducing noise and problems such as heteroskedasticity in the dataset (Arzandeh & Frank, 2019). Second, high-speed trading produces high-frequency market data. In order to understand trading actions, it is beneficial to observe the same data granularity as the trading algorithms. Such nanosecond LOB data is highly granular input data. If this data is collapsed/aggregated into snapshots, much information is washed out, and analyses will suffer.

In this paper, every single message is used – i.e. no snapshots were taken[4], except for comparison purposes. The analysis can handle message resolutions as detailed as those at the exchange itself, without the need for any aggregation. This means that there has been no further aggregation beyond the time precision and aggregation inherent in the CME data. Hence, all available information is used. Contrary to analyses that rely on the interpretation of snapshots, the use of highly granular data with irregular timing still allows for interpolations.

## 3. Research Design: CME LOB Data

Data consists of CME Group's proprietary market-depth dataset for all of 2015 for the U.S. Treasury Bond (T-Bond), corn, E-mini Dow Jones, crude oil, Henry Hub Natural Gas, soybeans, Chicago SRW wheat and rough rice futures markets. The files are in the CME Market Depth 3.0 (MDP) format, which provides the market messages required to recreate the LOB with millisecond precision. A market message is a set of tags and values which stores the data and metadata necessary to reconstruct the LOB. It is a sequential list of information about the LOB level without relation to other levels; this means that the preceding messages are needed to reconstruct the LOB (CME Group, 2020b).

---

[4] Note that LOB reconstructions based on message data also use snapshots, in that the LOB is updated whenever a new message is received. These snapshots are based on irregular time intervals, however, contrary to the time-based snapshots, which use regular time intervals. Throughout the paper, we use the term "snapshot" to refer to these time-based snapshots, as opposed to message-based snapshots.

Each file contains all contracts of the same futures contract ordered by message number and time of arrival. The LOB information is documented according to the FIX protocol (CME Group, 2020). This protocol uses incremental updates of a data sequence to share exchange data with traders and regulators, either live or in batches. It does not store the LOB itself, but only the messages that can be used to recreate the LOB (FIXtrading, 2020).

MDP data provides information about LOB levels, for example when a new price level is inserted in the LOB, the price or volume is changed at a particular level, or a price level is deleted from the LOB. Among other data, messages contain information about prices, bids, asks, trading/order quantities and the time of placing each limit and market order is placed in the platform, up to ten orders deep. Appendix B contains a description of the message types in each file and how the LOB is reconstructed.

Table I provides some descriptive statistics regarding the number of messages per day, messages per second, the total volume in the LOB per message, the total LOB value per message and the total number of messages for the contract duration of the December T-Bond, corn, E-mini Dow Jones and crude oil futures contracts in 2015.

**TABLE I**   Descriptive Statistics of the December T-Bond, Corn, E-mini Dow Jones and Crude Oil futures markets in 2015

|  | December T-Bond | December Corn | December E-mini Dow Jones | December Crude Oil |
|---|---|---|---|---|
| Time Window | May 1 – Dec 21 | Jan 1 – Dec 14 | Jan 1 – Dec 17 | Jan 1 – Nov 26 |
| Total # Messages | 171,330,999 | 112,378,245 | 200,514,606 | 520,742,261 |
| Messages per day | | | | |
| Minimum | 3 | 30 | 10 | 431,574 |
| Mean | 1,044,701 | 462,462 | 815,100 | 2,283,957 |
| Median | 407,078 | 425,095 | 43,622 | 2,073,330 |
| Maximum | 3,816,504 | 1,708,758 | 4,480,008 | 11,572,077 |
| Messages per second | | | | |
| Minimum | 0 | 0 | 0 | 0 |
| Mean | 13 | 7 | 10 | 28 |
| Median | 0 | 0 | 0 | 2 |
| Maximum | 4,295 | 14,958 | 5,475 | 4,105 |
| Total LOB Volume per Message | | | | |
| Minimum | 20 | 44 | 0 | 0 |
| Mean | 4,529 | 3,356 | 477 | 247 |
| Median | 4,729 | 2,859 | 519 | 136 |
| Maximum | 13,787 | 26,508 | 1,392 | 2,427 |
| Total underlying LOB value ($) per message | | | | |
| Minimum | 2,963,906 | 822,188 | 0 | 0 |
| Mean | 705,829,909 | 65,422,050 | 40,895,393 | 12,071,469 |
| Median | 737,803,000 | 56,137,338 | 43,314,230 | 7,436,740 |
| Maximum | 2,157,002,250 | 497,274,150 | 123,262,490 | 110,284,130 |

*Note:* Numbers are rounded. Each contract has a different time window because of different start and expiration dates. The total LOB value per message (in dollars) is the total sum of price levels multiplied by their respective volume, converted to dollars (e.g. T-Bonds trade in points and corn in dollar cents). Trading days where most CME futures markets were closed were taken out of the sample: New Year's Day (January 1), Good Friday (April 3), the day before Independence Day (July 3) and Christmas (December 25). Zeros in the table may be due to the recent launch of the contract or the contract nearing its expiry date.

## 4. Particle Physics Visualization of High-Frequency Data: CERN's ROOT

### 4.1. ROOT: Data storage

The software framework ROOT (Brun & Rademakers, 1997; CERN, 2018b) is used to reconstruct the LOB and create the visualizations. ROOT is developed by CERN, in conjunction with other parties, and is used to analyze large amounts of data, especially in particle physics. It is mainly written in C++ but is integrated with other languages such as Python, R and Mathematica (CERN, 2018b). All experiments at the LHC use ROOT to store and analyze their data. It is built to store large amounts of data – to date, the LHC project has stored more than one exabyte of data, i.e. 1,000,000 gigabytes – and process that data efficiently in a distributed setup (Tejedor & Kothuri, 2018). ROOT's strength lies in its bundling with Cling, a unique interactive C++ interpreter based on Clang and LLVM libraries (LLVM, 2021). This allows ROOT access to the best of both worlds: rapid development of code and more compiled optimizations for fast code execution. ROOT can be used to save, access and mine data, produce graphs, and it can run interactively or be used to build stand-alone applications (CERN, 2018a). Its key features include advanced data structures; reading and writing objects; graphics and visualization toolkit; and analysis modules[5].

A ROOT file (TFile) is a data container to store and bundle different types of related data, such as raw data, metadata or graphics. It allows for fast reading, compressed data storage and data access over networks. The data from CME Group is stored in ROOT files to facilitate analyses. While converting the data to ROOT, no reconstruction of the LOB is performed, since storing the full LOB – including every message – would require more storage space than reconstructing it on demand (the latter is done while generating the plot)[6]. The conversion to the ROOT file structure decreases the overall file size and increases the accessibility of the data. Subset data reconstructed in the ROOT format is about 15 times smaller than the corresponding raw data[7].

---

[5] A few examples of modules are the Toolkit for Multivariate Data Analysis with ROOT (TMVA) for multivariate machine learning, RooFit for data fitting and RooStats for statistical analysis. This allows users, among others, to model the expected distribution of events, use neural/deep networks, function discriminant analysis, support vector machines, multidimensional minimization, fitting, parametrization, and likelihood ratio tests for hypothesis testing (Antcheva et al., 2009; CERN, 2021a, 2021c, 2021b). Python and R bindings are automatically generated for the full framework, which reduces the learning curve and complexity of development. Furthermore, ROOT is backwards compatible which reduces challenges in using old code.

A caveat of ROOT is that it is primarily written for the needs of particle physics and, hence, might not offer tools that are considered basic in other disciplines. Also, ROOT is backward compatible, which has advantages (e.g. "old" code still works with the current ROOT version) as well as disadvantages in the form of odd design choices and practices. For example, ROOT has its own classes (e.g. TList or THashList) where one would typically expect C++ Standard Template Library (STL) containers. This is because STL was still in its infancy when ROOT was developed and needed such classes. In addition, ROOT is not in any textbooks nor taught in computer/data science, which may act as a barrier to using it.

---

[6] The LOB reconstruction for a full year is relatively fast and complements previous research on processing LOB data (Gai et al., 2014).

[7] ROOT is built to work on a computing cluster, but the core program is relatively small and can run on almost any computer, with as little as 500MB of RAM. The algorithms developed for this research use a streaming architecture, meaning that the data is loaded in chunks, so that even terabytes of data can be processed on a computer with just 500MB of RAM. This also means that the algorithms are not limited by file size, since only a part of the data is loaded into memory at any point in time. The time complexity of the algorithm is linear towards the number of messages (O(n)). Code can reconstruct the LOB from message data at a rate of up to 180MB per second, or, depending on additional processing steps, between 200,000 to 500,000 messages per second.

## 4.2. Visualization in ROOT

As particle physics analysis mostly involves statistical distributions, one of ROOT's main features for aggregation and visualization is the use of histograms. Histograms offer a natural technique of reducing data and visualizing distributions, summarizing millions of single measurements of a quantity under study by filling a few bins (i.e. histogram bars). Consequently, ROOT offers various histogram classes, such as TH1D, a one-dimensional histogram storing double-precision floating-point numbers, TH2D, its two-dimensional counterpart; as well as three- and N-dimensional histograms (TH3D, THn). The x-axis of one-dimensional histograms aggregates different value ranges of a quantity, while the y-axis shows event counts or densities for each range (Antcheva et al., 2009).

Histograms are not only used to count integers, but also to weight entries in the datasets. This is often necessary when analyzing simulations, since the probability of detecting a specific simulated event has to be taken into account to generate the correct frequency distributions. For example, traded volume can be visualized, where each trade fills a histogram bin and is weighted by the trade's volume. Hence, the total content of the bin is the sum of all trades' volumes for a given parameter range. ROOT histograms automatically compute the statistical (Poisson) uncertainty in each bin and take weights into account (Antcheva et al., 2009). The data in histograms can be further manipulated: histograms can, for example, be re-binned, smoothed, added, scaled, subtracted or projected to analyze their content. In finance, the use of binned axes – especially for the abscissa – is uncommon. In this paper, histograms are used to plot messages, but the aggregation tool is not used to visualize the reconstructed LOB: since each bin corresponds exactly to one message, there is no counting involved. In the next sections, we show the value of data visualization using ROOT in the high-frequency finance context of the LOB.

ROOT's graphical interface is able to plot advanced graphs and histograms. ROOT can display multiple panels in one window, each providing a coordinate system to plot objects. Since a window can host multiple panels, different quantities can be visualized simultaneously. In this paper, for example, a panel showing a two-dimensional weighted histogram is displayed together with one or more panels showing one-dimensional histograms, where all panels have identical x-axes but different y-axes. With this technique, for example, the evolution of many variables versus time can be displayed and interactively manipulated[8].

## 5. Particle Physics Visualization of the LOB

In this study, a two-dimensional histogram is used to visualize the state of the LOB. We illustrate how the visualization can be adapted to preferences and research topics of interest. The dates and time windows visualized are selected for illustrative (visualization) purposes only; these visualizations are not meant to lead to any conclusions. Example code of the visualizations and online, zoomable versions of all the visualizations are made public in a GitHub repository (https://github.com/HighLO).

---

[8] Generating the visualizations in this paper takes around 40 seconds on a standard laptop, with most of the computation power going to extracting the FIX messages and reconstructing the order book. The code developed for the proposed visualization methodology runs over the data twice: 1) once to extract the statistics needed for the visualization, such as the minimum and maximum price that are used on the y-axis of the LOB and 2) once more to reconstruct the LOB. Note that the code can be altered to only run over the code once (dynamically increasing/decreasing the maximum and minimum price), in which it takes just 20 seconds to generate the plots. It can reconstruct the LOB from message data at a rate of up to 180MB per second, or, depending on additional processing steps, between 200,000 to 500,000 messages per second.

Figure 1 shows the LOB of the T-Bond futures market on November 12, 2015 between 09:00 AM and 10:00 AM CT. The x-axis of the histogram shows the message sequence number within the time window studied. The y-axis shows the range of prices between which the LOB moves. The histogram is filled by reconstructing the LOB for every message. For each message, the bins are filled with the volume at each LOB price level. This means that every bin is only filled once – contrary to the traditional, aggregating use of histograms. The midpoint, i.e. the middle of the first bid and first ask price, is visualized by a red line. The ask levels are above and the bid levels below the red line. In addition to the LOB histogram, the cumulative traded volume is shown at the beginning of the time window in a separate panel for every message. The visualization uses messages; however, time is also of importance as there are time-priority rules in place prioritizing orders that arrive first (Yao & Ye, 2018). Therefore, the progression of time accumulated since the beginning of the time window is shown in the lower panel of the figure for every message. A steeper (flatter) line signals a lower (higher) rate of messages and, thus, a lower (higher) level of market activity. In other words: a steeper (flatter) line signals more (less) time progression. Hence, our visualizations capitalize on both messages (market activity) as well as snapshots (time), in that both are visualized simultaneously. All panels are combined into one graph with the same x-axis, i.e. message number. The visualization can also be performed using snapshots instead of messages, i.e. with time on the x-axis. If, for example, the snapshot size is set at one second, the histogram only plots the LOB of the last message of each second.

Data of the top 10 levels of the LOB was recorded in 2015. Hence, a total of 20 levels are plotted vertically for any message: the top 10 levels in the figures show the ask side and the bottom 10 levels show the bid side. Data outside these ranges still exists at the exchange, but traders did not see these levels and the exchange did not emit any messages for these price levels. Note that sometimes the last levels on both the bid and ask side are not completely filled. For example, the first and second levels may disappear, as either the volume at these levels was cancelled or a trade took place that consumed the volume of these levels. This would leave the LOB empty on the ninth and/or tenth level for a particular message.

Figure 1 shows that, on November 12 between 09:00 AM and 10:00 AM CT, the exchange received approximately 215,000 messages for the T-Bond futures market. The top panel shows that the LOB remained relatively stable, fluctuating between 152.1 and 153 points. Colors indicate the volume of the price level. For example, at the 152.75 points level, there was a consistently high volume throughout the time window, as indicated by the yellow color. This means that many sell orders were resting at this price level, waiting to be executed as soon as the price would reach this level. Cumulative trade volume rose at a steady pace (meaning that trades took place consistently), with a few spikes, e.g. at the 95,000, 135,000 and 140,000 message marks. Time progressed relatively stably, meaning that messages kept arriving at a regular rate, i.e. there were no periods with more or fewer messages arriving – which would have indicated more or less market activity.

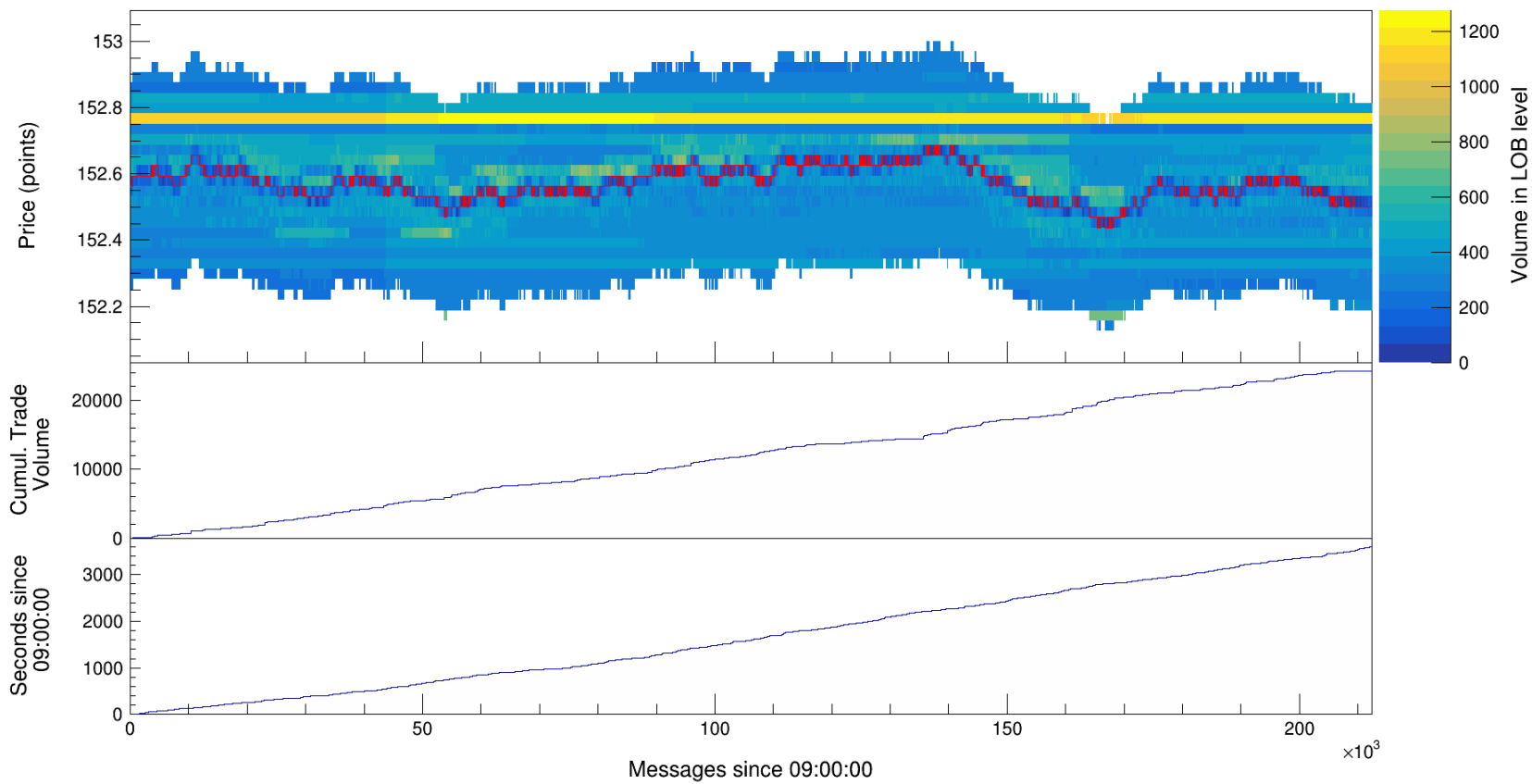

**Figure 1** December U.S. Treasury Bond (ZBZ5) LOB behavior (November 12, 2015 from 09:00 AM to 10:00 AM CT). The top panel shows the volumes at the 10 price levels on the bid and ask side of the LOB, respectively. Each unit on the x-axis is one message. The y-axis represents the price of the T-Bond in points. The color of each bin represents the volume in the LOB at that message and price. The scale ranges from blue to yellow, with the color becoming a brighter yellow as volume increases at that price level. The red line is the midpoint. The middle panel shows the cumulative trading volume for the selected time horizon. A steeper (flatter) line signals a higher (lower) rate of traded volume. The bottom panel shows how much time passes between messages reported by the exchange. A steeper (flatter) line signals a lower (higher) rate of messages, given that a steeper (flatter) line signals more (less) time progression. (data source: author's visualization of CME MDP 3.0 Market Data).

In the sections below, the visualizations are modified to illustrate the benefits of using particle physics visualizations on LOB data. First, we illustrate the added value of using message numbers rather than time (i.e. snapshots) on the x-axis by visualizing the LOB of the December Corn contract for a U.S. Department of Agriculture (USDA) announcement day (Wednesday August 12, 2015). Second, we illustrate various ways to visualize trade volume and time by way of the December E-Mini Dow Jones futures contract. Third, two related markets – the December crude oil and December Henry Hub natural gas futures contracts – are plotted using the same timeframe. Fourth, the proposed methodology can help to visualize other variables as is demonstrated for liquidity in Section 6. Finally, we show an additional visualization possibility by ROOT that compresses large amounts of data in a single figure in Section 7.

### 5.1. Snapshot vs. Message

This section visualizes the LOB of the December Corn futures contract on a USDA announcement day, to illustrate the difference between using snapshots and messages in visualizations. First, Figure 2 shows a histogram of the December Corn futures contract to give an impression of how many messages arrive within one second. The x-axis shows the number of messages within one second; the y-axis is a logarithmic scale and shows how often a 'number of messages within one second' occurs in the dataset; each bar (or step/bin) represents 20 messages. Figure 2 shows that there are over ten million seconds that contain between 0-20 messages each. Next, there are more than 250,000 seconds that contain 20-40 messages each. The frequency of seconds decreases steadily as the number of messages per second increases, reaching a low of approximately 8 seconds that include 2000 messages each. Figure 2 thus illustrates that using one-second snapshots typically leads to a loss of information, as one second frequently contains multiple messages.

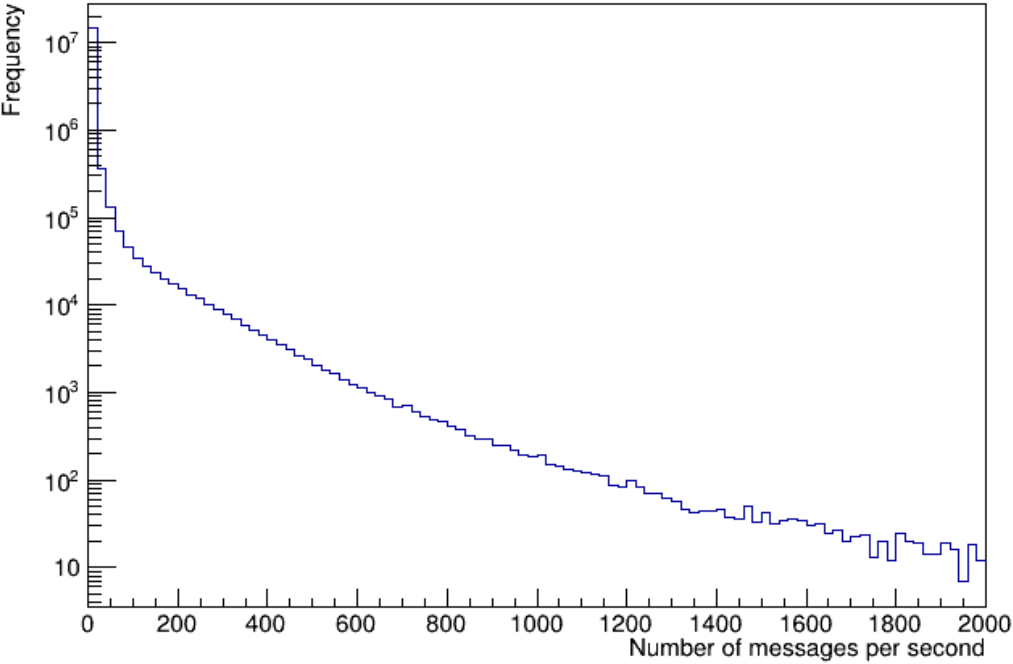

Figure 2    Histogram of number of messages per second for the December Corn futures contract data in 2015 (ZCZ5). (data source: author's visualization of CME MDP 3.0 Market Data).

Next, the December Corn futures contract is visualized for a three-hour window on a USDA announcement day – August 12, 2015 from 09:00 AM to 12:00 AM CT – using both snapshots and messages. The announcement forecasted corn production to be 156 million bushels higher than projected in July. Supplies were forecasted to be 154 million bushels higher than projected in July, reaching a record of 15.5 billion bushels.

Figure 3 shows the LOB when using snapshots of five seconds and Figure 4 when using messages. The five-second snapshot was chosen arbitrarily for illustrative purposes. Figure 3 shows a steep drop at approximately 7000 seconds after 09:00 AM CT, exactly after the USDA report was announced, at 11:00 AM CT. The same drop can be seen in Figure 4 at approximately 210,000 messages after 09:00 AM CT. Simultaneously, trading increased significantly, as can be seen by the steep rise in cumulative trade volume in the second panel. Figure 3 shows two large LOB price drops: one immediately after the USDA announcement (at approximately 7000 seconds) and another one after the slight increase following the first drop (at approximately 7800 seconds). However, a different pattern emerges when visualizing all of the messages, as per Figure 4, rather than using snapshots. Instead of an immediate steep drop after the announcement, the LOB decreases (around the 210,000 message mark), after which it immediately recovers (around the 215,000 message mark) only to decrease further at a steady pace (around the 225,000 and the 305,000 message marks). Hence, Figure 4 shows three LOB price drops rather than the two drops identified in Figure 3. In addition, the increases and decreases after the steep LOB price drop are better visible in the message-based figure than in the snapshot-based figure. This is not due to zooming but to messages being compressed into seconds to create snapshots and, thus, to loss of information. The two figures illustrate that snapshots cannot reveal all of the information and market activity in the LOB: patterns can be better studied, in more detail, and visualizations can incorporate more information when using messages rather than when using snapshots (for example, orders that are added and subsequently almost immediately cancelled are not observable via snapshots but do become visible when using messages). Figure 4 also constitutes a novel method to visualize messages and time simultaneously, thus fully retaining any time-related information.

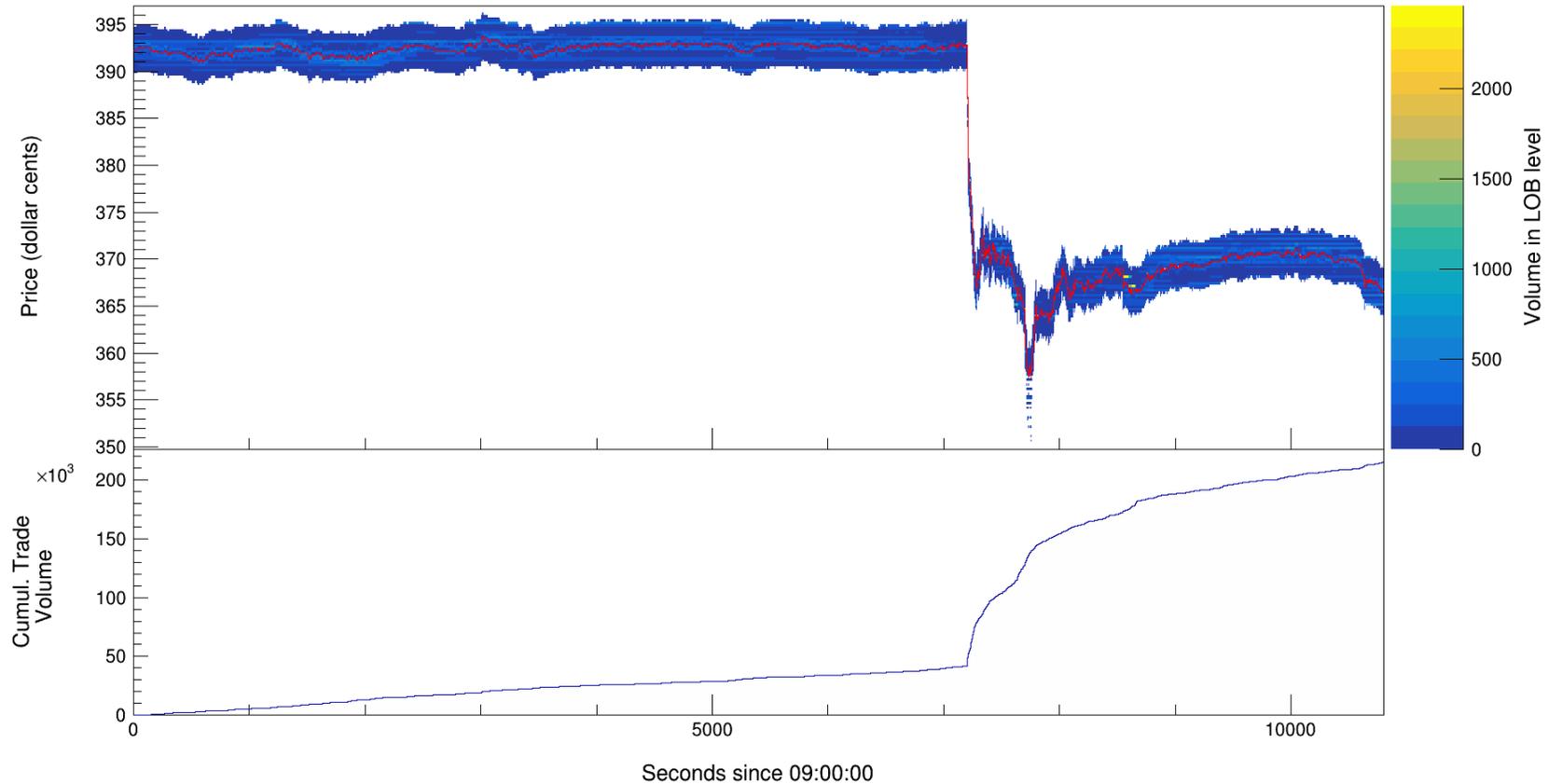

**Figure 3**   December corn (ZCZ5) LOB behavior on an USDA announcement day using 5-second snapshots (August 12, 2015 from 09:00 AM to 12:00 AM CT). The top panel shows the volumes at the 10 price levels on the bid and ask side of the LOB, respectively. Each unit on the x-axis is one second. The y-axis represents the price of corn in dollar cents. The color of each bin represents the volume in the LOB at that second and price. The scale ranges from blue to yellow, with the color becoming a brighter yellow as volume increases at that price level. The red line is the midpoint. The bottom panel shows the cumulative trading volume per second. A steeper (flatter) line signals a higher (lower) rate of traded volume. (data source: author's visualization of CME MDP 3.0 Market Data).

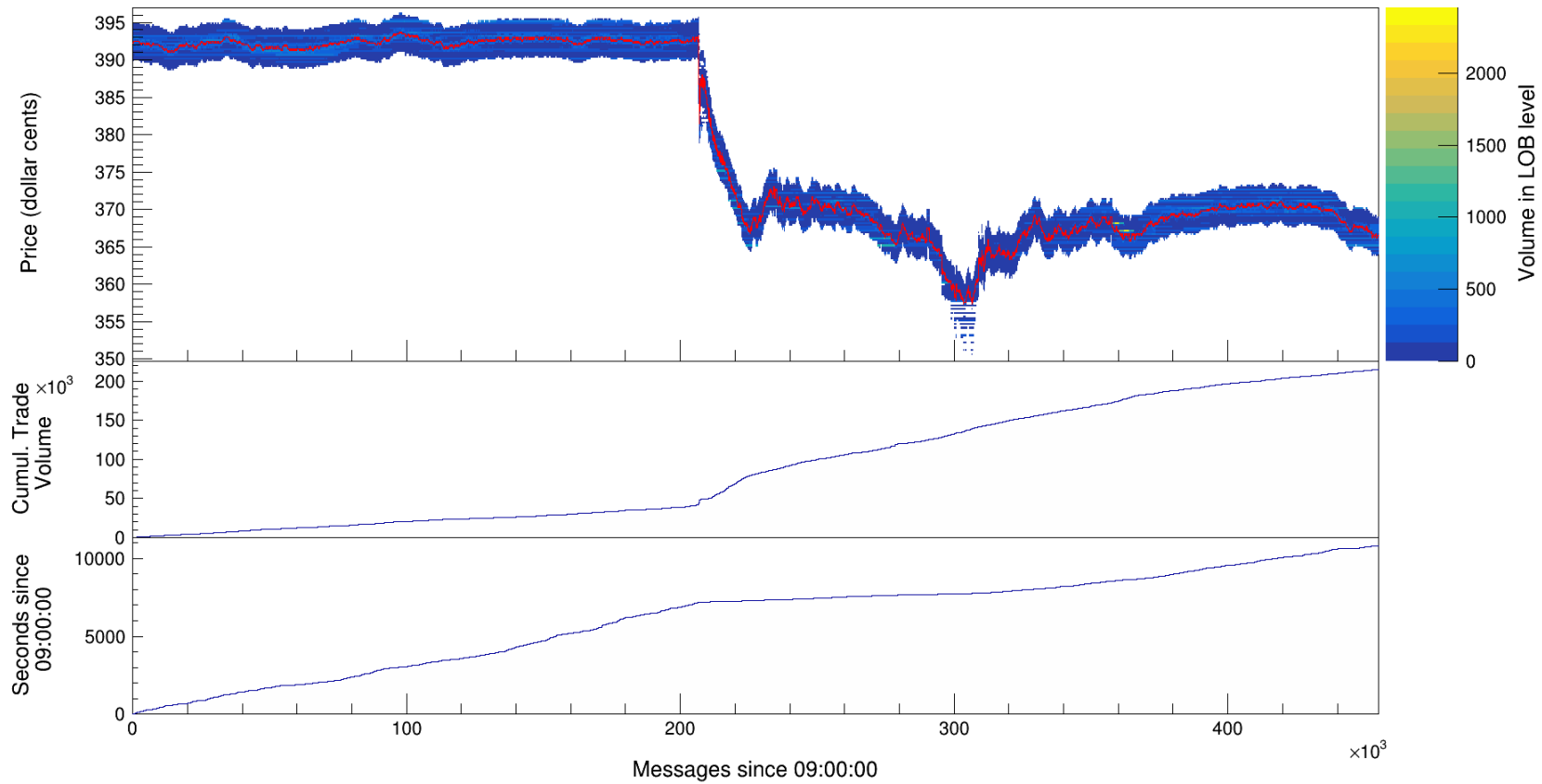

**Figure 4** December corn (ZCZ5) LOB behavior on an USDA announcement day using messages (August 12, 2015 from 09:00 AM to 12:00 AM CT). The top panel shows the volumes at the 10 price levels on the bid and ask side of the LOB, respectively. Each unit on the x-axis is one message. The y-axis represents the price of corn in dollar cents. The color of each bin represents the volume in the LOB at that message and price. The scale ranges from blue to yellow, with the color becoming a brighter yellow as volume increases at that price level. The red line is the midpoint. The middle panel shows the cumulative trading volume for the selected time horizon. A steeper (flatter) line signals a higher (lower) rate of traded volume. The bottom panel shows how much time passes between messages reported by the exchange. A steeper (flatter) line signals a lower (higher) rate of messages, given that a steeper (flatter) line signals more (less) time progression. (data source: author's visualization of CME MDP 3.0 Market Data).

## 5.2. Trade Volume and Time

The visualization can be modified to display many variables in various ways. Figure 5 illustrates this for trade volume and time by way of the December E-mini Dow Jones futures contract. In addition to cumulative trade volume, the middle panel shows trade volume per five-second snapshot in green bars. Note that this snapshot size was set by the authors. It is also possible to visualize the number of trades per message – though this would be less easily interpretable as there are more than 370,000 messages and, thus over 370,000 bars would have to be visualized. Around the 365,000 message mark, trade volume per snapshot rose to approximately 300 contracts. The LOB responded by increasing around the same message mark. This increase in trade volume is less discernable when looking at cumulative trade volume. Visualizing the same variable (i.e. trade volume) in various ways can thus result in different insights. In addition to seconds (since the start of the time window), time is visualized in 'messages per five-second snapshot'. This method of visualizing time gives a better impression when the exchange reports many messages. For example, around the 135,000 message mark, more messages occur within several snapshots, indicating more activity in the LOB. This is less visible in the visualization of the T-Bond market (Figure 1), which visualizes the number of seconds (since the start) of the plot. Depending on one's goals, variables can thus be visualized in various ways to study them from different perspectives and to gain more knowledge about LOB behavior.

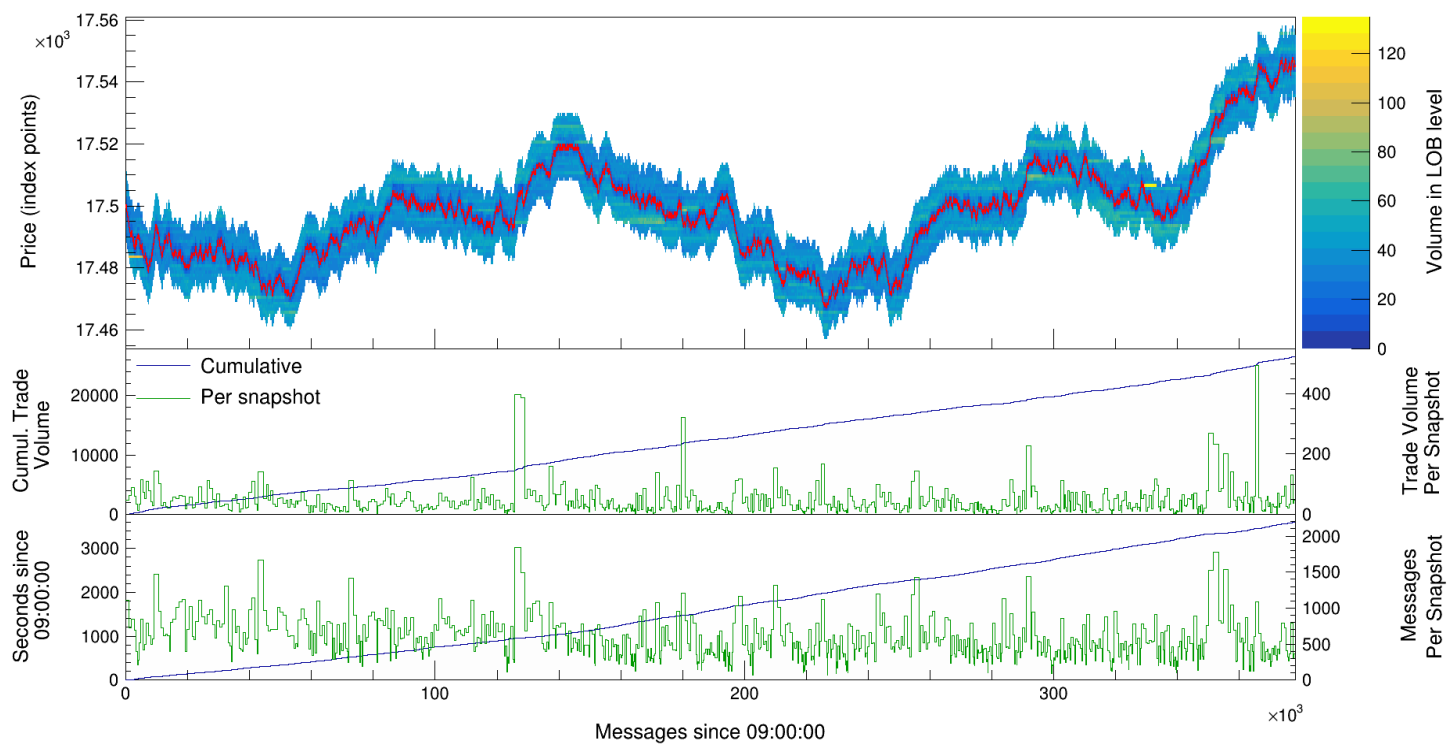

**Figure 5**   December E-mini Dow Jones (YMZ5) LOB behavior (November 12, 2015 from 09:00 AM to 10:00 AM CT). The top panel shows the volumes at the 10 price levels on the bid and ask side of the LOB, respectively. Each unit on the x-axis is one message. The y-axis represents the price of the E-mini Dow Jones in index points. The color of each bin represents the volume in the LOB at that message and price. The scale ranges from blue to yellow, with the color becoming a brighter yellow as volume increases at that price level. The red line is the midpoint. The middle panel shows the cumulative trading volume for the selected time horizon and the trade volume per snapshot (5 seconds). The blue line indicates the cumulative trade volume on to the left y-axis; a steeper (flatter) line signals a higher (lower) rate of traded volume. The green bars show the total traded volume in a 5-second window on the right y-axis. The bottom panel shows how much time passes between messages reported by the exchange and how many messages occur within one snapshot of 5 seconds. A steeper (flatter) blue line signals a lower (higher) rate of messages, given that a steeper (flatter) blue line signals more (less) time progression. The blue line is related to the left y-axis. The green bars are related to the right y-axis and show the number of messages that occur within one snapshot of 5 seconds. (data source: author's visualization of CME MDP 3.0 Market Data).

### 5.3. Two Markets in One Visualization

The visualization can also be modified to show two (or more) LOBs simultaneously. However, careful attention must to be paid to how the LOBs are synchronized, as message numbers are not necessarily aligned between two markets. For example, a market with little activity may have 100 messages until 10:00 AM, whereas an active market may have 1000 messages in the same time window. To solve this, all messages are merged and ordered into a single time series. If more messages arrive in Market A, the LOB for Market B is repeated and not updated until a new message arrives for Market B. In other words, the last known LOB is retained until something changes (i.e. a new message arrives). Consider the following example of a merged list between Market A and Market B: $[A_1, B_1, A_2, A_3, B_2]$. The LOBs would be visualized as per Table II.

**TABLE II**      Example of merging two LOBs for visualization purposes

| **$A_1$** | $A_1$ | **$A_2$** | **$A_3$** | $A_3$ |
|---|---|---|---|---|
| $B_0$ | **$B_1$** | $B_1$ | $B_1$ | **$B_2$** |

*Note:* Bold letters indicate the first time a message occurs.

Figure 6 illustrates the LOBs for the December crude oil (CLZ5) and December natural gas (NGZ5) futures contracts on November 12 between 09:00 AM and 10:00 AM CT. The top panel shows the LOB movements in both markets. More trading occurs in the crude oil futures market, with approximately 40,000 futures contracts being traded compared to less than 10,000 in the natural gas market. Furthermore, messages arrive at a regular pace, as shown by the steadily rising timeline in the bottom panel. Visualizing two LOBs within one graph can be beneficial when studying relationships between markets, the effect of an event on either market or the spillover effects from one market into another (e.g. regarding trades, volatility or liquidity).

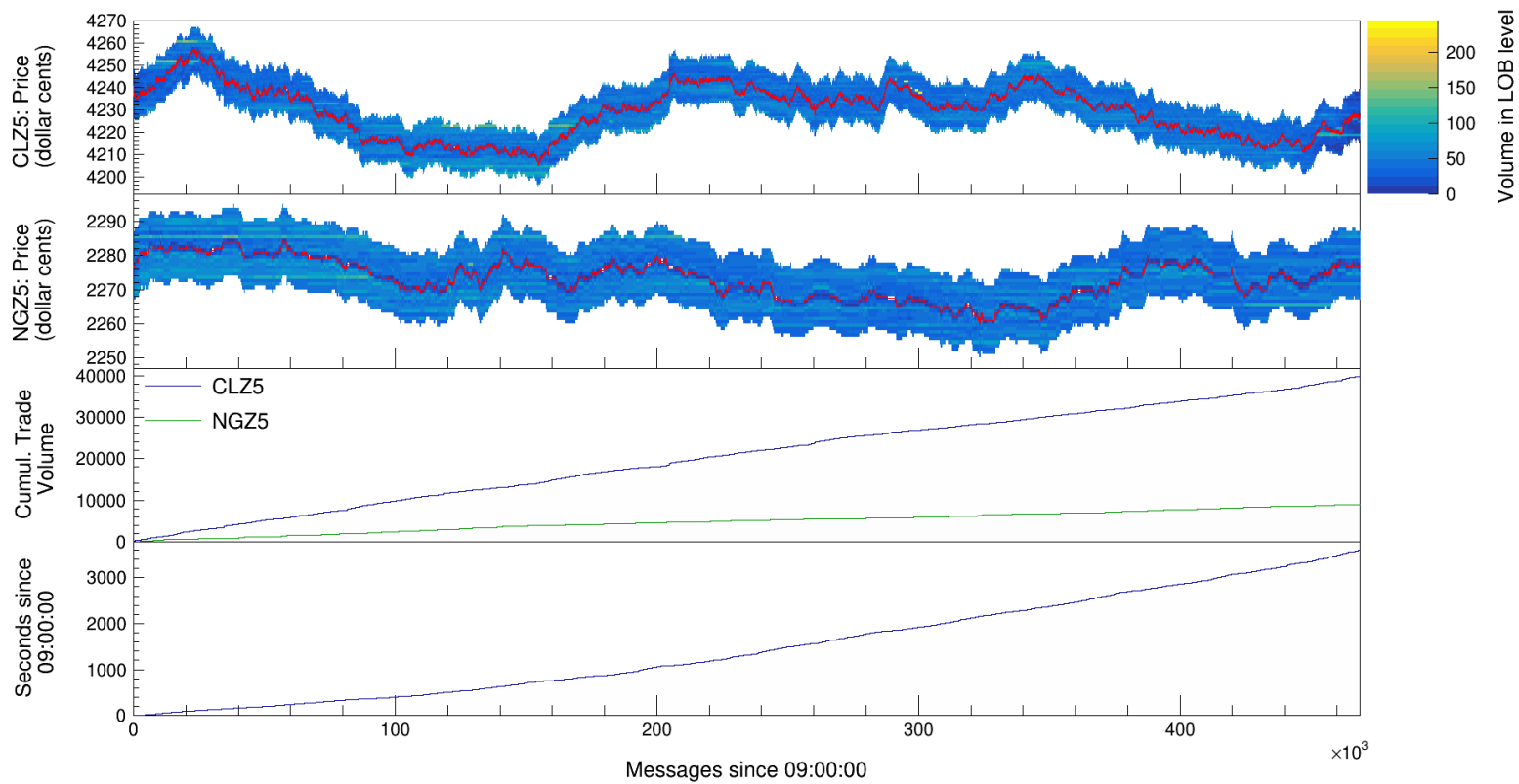

**Figure 6**    December Crude Oil (CLZ5) and December Henry Hub Natural Gas (NGZ5) LOB behavior (November 12, 2015 from 09:00 AM to 10:00 AM CT). The top two panels show the volumes at the 10 price levels on the bid and ask side of the LOB, respectively, for the December crude oil futures contract (CLZ5) and the December natural gas futures contract (NGZ5), respectively. Each unit on the x-axis is one message. The y-axes represent the respective prices of crude oil and natural gas in dollar cents. The color of each bin represents the volume in the LOB at that message and price. The scale ranges from blue to yellow, with the color becoming a brighter yellow as volume increases at that price level. The red line is the midpoint. The third panel shows the cumulative trading volume for the selected time horizon for both markets. The blue line indicates the cumulative trade volume for crude oil and the green line for natural gas. A steeper (flatter) line signals a higher (lower) rate of traded volume. The bottom panel shows how much time passes between messages reported by the exchange. A steeper (flatter) line signals a lower (higher) rate of messages, given that a steeper (flatter) line signals more (less) time progression. (data source: author's visualization of CME MDP 3.0 Market Data).

# 6. Particle Physics Illustration of the LOB: Illustration for Liquidity

The proposed visualization methodology can be adapted to many research topics. It helps answering research questions concerning, for example, market microstructure and optimal trading frequency (Du & Zhu, 2017). In this paper, it will be illustrated by way of liquidity, an important factor for the functioning of financial markets and one of the major topics in financial research (Capponi, Menkveld, & Zhang, 2019; Kerr, Sadka, & Sadka, 2019; Li, Wang, & Ye, 2019; Menkveld & Zoican, 2017; Peress & Schmidt, 2020; Trebbi & Xiao, 2019). Liquidity is illustrated since it is a concept with multiple dimensions, which makes it relatively complex to visualize in a single visualization. Liquidity consists of four dimensions: immediacy, tightness, depth and resiliency (Hasbrouck, 2017; Kyle, 1985). Most LOB visualizations in the literature show (some form of) liquidity, i.e. the volume in the market at different price levels ("depth") and the bid-ask spread ("tightness").

Aidov & Daigler (2015) visualize the LOB in two ways. Figure 1 in their paper (column 2 in Table III) visualizes the cumulative depth for each level across the book over time, i.e. the bid and ask volumes aggregated per level. It gives a better insight into how total volume behaves at each level and how it is distributed across time. However, the bid and ask side can have asymmetrical depths. By visualizing their cumulative depth, information might get lost, as it becomes impossible to disentangle the individual behaviors of the bid and ask side. In addition, no price levels are included. This makes it challenging to examine certain price-related volume patterns in the LOB, such as shifts in depth related to trades or a volatile market. Figure 2 in their paper (column 3 in Table III) visualizes the shape of the LOB on one day for five levels on the bid and ask side. The visualization shows a quick and easy-to-understand snapshot of the LOB. However, averages of averages are taken (averaging depth across 5-minute intervals which are then averaged over the day), and it shows a "snapshot" of one day rather than the distribution over time. Hence, a lot of LOB information and behavior is lost. Paddrik et al. (2016) propose many visualizations for regulators to use, two of which are specific to the LOB. Figure 7 in their paper (column 4 in Table III) visualizes the LOB with a heatmap of the depth at various price levels, i.e. different colors are used to indicate volume on the bid and ask side. It presents an instant picture of LOB behavior over time and the colors offer a quick overview of distributed volume. Figure 8 in their paper (column 5 in Table III) represents a simultaneously animated visualization of the LOB and stop loss order book that includes many textual variables. Since trading takes place in nanoseconds, the animated information might change fast, making it less suitable for real-time animations and quick decision-making by market participants and regulators.

All visualizations would benefit from the inclusion of trades, as trades contain information, take liquidity from the LOB and can explain certain behavior. Informed traders, for example, prefer larger trades (Easley & O'Hara, 1987). This helps us understand LOB movements; for example, the LOB might respond to large trades, which would remain unobserved in any of the figures included in Table III. Furthermore, all of these visualizations use time (i.e. snapshots), the disadvantages of which were discussed in Section 2, rather than messages. While Figure 7 from Paddrik et al., (2016) shows the LOB in a heatmap format, in this paper, it is extended by the ability to render LOB data with user specifications, such that multiple variables can be added and visualized.

TABLE III   Comparison between various LOB visualizations

| Authors | Aidov & Daigler (2015) | | Paddrik et al. (2016) | | Proposed Visualization |
|---|---|---|---|---|---|
| Figure # in article | 1 | 2 | 7 | 8 | 6 and 7 |
| Message/Time | Time | Time | Time | Time | Message |
| Time interval | 5 minutes | Daily | 100ms | ms | ms |
| Variables | Time, Depth, LOB level | Day, Bid and ask percentage quote, Bid and ask depth, Day and Night | Time, Price, Volume (per level per side), Bid and Ask | Time, Price stop loss and limit order book, Depth stop loss and limit order book, Bid and Ask side, Best bid, Best ask, (Cumulative) Volume, Most new order†, Most canceled†, Most modified†, Most aggressive†, Most passive† | Message number, Price, Volume (per level per side), Midpoint, Cumulative trade volume, Bid volume, Ask volume, Time |
| Volume | Bid and ask aggregated per level | Mean depth per day | Per level per side | Per level per side and cumulative volume | Per level per side, cumulative volume and total volume per side |
| Price | No | Relative | Absolute | Absolute and best bid and best ask | Absolute |
| Trades | No | No | No | No | Yes |
| Bid-ask spread | No | Relative | Can be deduced | Yes | Can be deduced or added |
| Midpoint | No | Relative | Can be deduced | Can be deduced | Yes |

† Includes participant identification number and number of contracts.

The methodology proposed in this paper complements the existing liquidity visualization tools. It visualizes all dimensions of liquidity and can easily be modified to display different and/or more variables. To demonstrate the relevance and power of the visualization, a liquid market – the Chicago SRW Wheat futures contract – is compared to a less liquid market, – the Rough Rice futures contract – on October 13, 2015, 9:00 AM – 12:00 AM CT. Figure 7 shows the Chicago SRW Wheat futures market, which is relatively liquid compared to the rough rice futures market, displayed in Figure 8. The visualization as illustrated in Section 5 is modified to include more liquidity variables: total volume on the bid (yellow line) and ask (pink line) side. Other liquidity variables, such as the bid-ask spread or volume per level, can also be visualized.

With more than 180,000 messages compared to 4500 in the rough rice market, the wheat market shows more activity within the same time frame. The LOB moves smoothly and is compact in the wheat market, i.e. all orders in the LOB rest in adjacent price levels. In contrast, the rough rice market in Figure 8 shows frequent gaps – indicated by white spaces – where there is no volume for that specific price level. The wheat market displays a steady and smooth increase of cumulative trade volume and time progression, contrary to the rough rice market, where cumulative trade volume and time are subject to large spikes. This means that messages arrive more irregularly in the rough rice market – and more regularly in the wheat market. As indicated by the flat line, the time between messages is shorter towards the end of the time window in the rough rice market. This indicates a higher frequency of messages and, thus, more activity in the LOB than before. This illustrates once again the advantage of using messages over snapshots: to create snapshots, messages and information are collapsed into regular time intervals, which might have led to different conclusions for the rough rice market. Therefore, the LOB behavior as revealed in Figure 8 might not have been observable by way of snapshots.

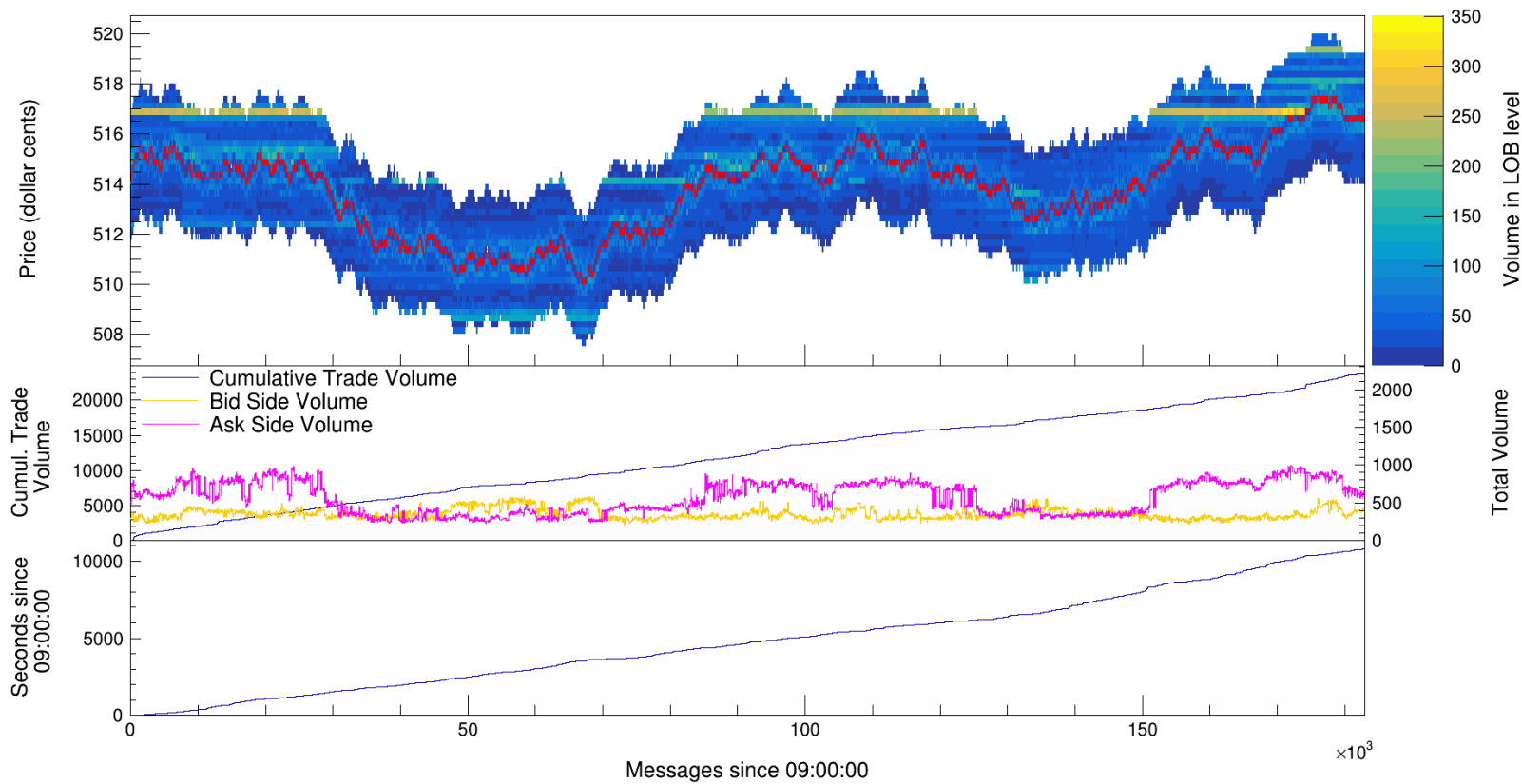

**Figure 7** December Chicago SRW Wheat (ZWZ5) LOB behavior (October 13, 2015 from 09:00 AM to 12:00 AM CT). The top panel shows the volume at the 10 price levels on the bid and ask side of the LOB, respectively. Each unit on the x-axis is one message. The y-axis represents the price of SRW wheat in dollar cents. The color of each bin represents the volume in the LOB at that message and price. The scale ranges from blue to yellow, with the color becoming a brighter yellow as volume increases at that price level. The red line is the midpoint. The middle panel shows the cumulative trading volume for the selected time horizon and the total volume on the separate sides of the LOB. The blue line indicates the cumulative trade volume on the left y-axis. A steeper (flatter) line signals a higher (lower) rate of traded volume. The pink line represents the total volume on the ask side and the yellow line the total volume on the bid side on the right y-axis. The bottom panel shows how much time passes between messages reported by the exchange. A steeper (flatter) line signals a lower (higher) rate of messages, given that a steeper (flatter) line signals more (less) time progression. (data source: author's visualization of CME MDP 3.0 Market Data).

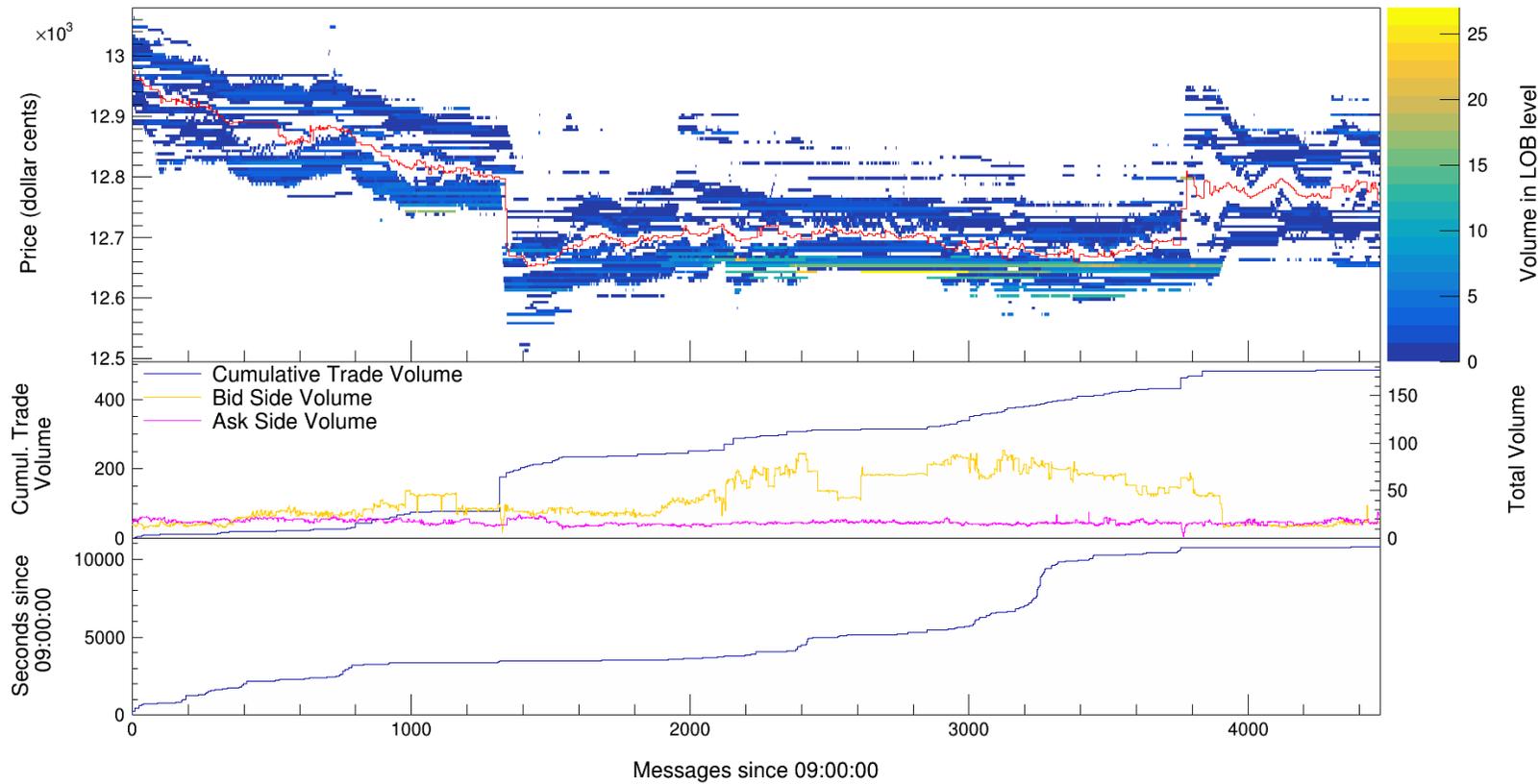

**Figure 8** December Rough Rice (ZRX5) LOB behavior (October 13, 2015 from 09:00 AM to 12:00 AM CT). The top panel shows the volume at the 10 price levels on the bid and ask side of the LOB, respectively. Each unit on the x-axis is one message. The y-axis represents the price of rough rice in dollar cents. The color of each bin represents the volume in the LOB at that message and price. The scale ranges from blue to yellow, with the color becoming a brighter yellow as volume increases at that price level. The red line is the midpoint. The middle panel shows the cumulative trading volume for the selected time horizon and the total volume on the separate sides of the LOB. The blue line indicates the cumulative trade volume on the left y-axis. A steeper (flatter) line signals a higher (lower) rate of traded volume. The pink line represents the total volume on the ask side and the yellow line the total volume on the bid side on the right y-axis. The bottom panel shows how much time passes between messages reported by the exchange. A steeper (flatter) line signals a lower (higher) rate of messages, given that a steeper (flatter) line signals more (less) time progression. (data source: author's visualization of CME MDP 3.0 Market Data).

The proposed visualization methodology complements the other visualization techniques discussed in Table III. All available information is embedded thanks to the use of messages rather than snapshots. It gives a quick overview of how volume is distributed, how active the market is, whether the bid-ask spread is tight or wide, what the midpoint price is and how trades affect each side of the market. Additional (liquidity) variables can be added to gain more insights.

7. Additional Visualization Possibilities with ROOT: an Example

To illustrate ROOT's additional visualization possibilities, a different and new visualization is shown that combines large amounts of data in a single figure. Figure 9 shows the development of trade prices before and after a transaction, for the December E-mini Dow Jones futures contract of 2015.

Figure 9 is a 2-dimensional histogram, displaying trade price behavior before and after a transaction. It is illustrated using two parameters, the trigger and the relative impact (though more parameters can be included). The trigger is the event under study, in this example transactions (other examples of triggers include cancellations in the LOB, liquidity measurements, USDA announcements, news articles or the opening of the market, etc.). Transactions (triggers) are set at time zero on the x-axis. The price of a transaction is located on the zero on the y-axis. All trades that take place in the specified time window are relative to the price level of the transaction (relative impact). For every transaction in the dataset, a time window is filtered of 50ms before and 50ms after the transaction. All trades that occur within this time window are counted for and put in their respective bins. Each bin represents the count frequency of timestamp-price combinations. ROOT counts the number of times that certain trade timestamp-price combinations occurs. The color of each bin represents the number count assigned to that bin. The logarithmic scale on the z-axis ranges from blue to yellow, with the color becoming a brighter yellow as the count increases.

To illustrate, 50ms after a transaction, approximately one million times a trade took place at 1 index point (E-mini Dow Jones) above the transaction. Figure 9 is a cumulative visualization, which allows for large datasets to be captured in a single plot: here it contains a total of 7,621,048 transactions. While many more inferences can be made from this figure, this is beyond the scope of this paper.

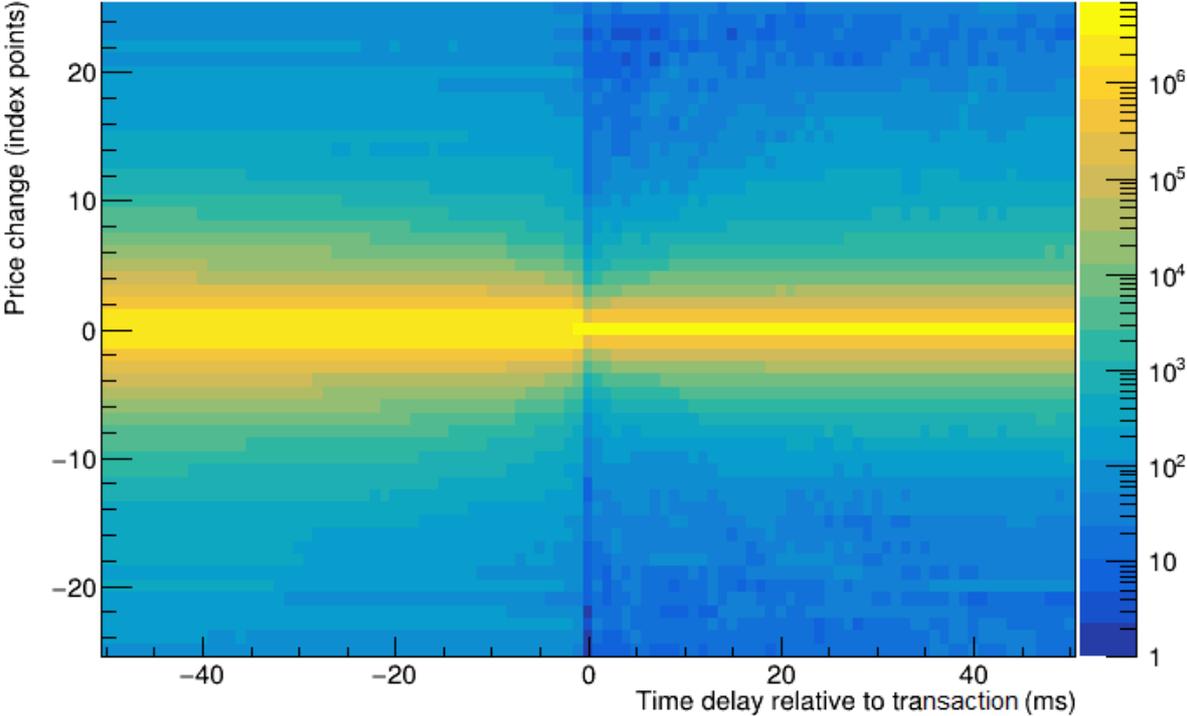

**Figure 9** Evolution of trade prices before and after a transaction for the December E-mini Dow Jones futures contract of 2015 (YMZ5). (data source: author's visualization of CME MDP 3.0 Market Data).

## 8. Discussion and Conclusion

The growing availability of LOB data poses new challenges to academics, industry, and regulators. LOB data is complex and can be difficult to reconstruct, visualize and analyze as it is irregularly spaced, high-speed and voluminous. This paper introduces a new method to visualize the LOB by looking at data through a particle-physics lens, which enables a better understanding of markets and the behavior in those markets, given that it takes all activities (messages) and preserves all of the information embedded in market participants' actions.

The proposed methodology to visualize the LOB complements the existing methods. While the existing methods are limited as to the number of variables they can display, the visualization proposed in this paper is richer and more easily adjustable with user specifications to many different types of research. It complements the previous literature in the following ways. First, all available messages are used and no snapshots are taken. This reveals the actual changes in the LOB, so that patterns in the LOB can be studied better. This is especially interesting when studying the behavior of HFTs, where the use of snapshots will make their behavior more opaque and harder to study, given that they trade within nanoseconds. Second, since the visualizations are based on messages, they are no longer limited by and dependent on time. Thus, given that the time variable is no longer the constraining factor, visualizations can be created that would otherwise have been impossible, and more relevant variables and their relationships can be shown. Third, the visualization shows the changing distribution of the LOB over time, including the trades that take place. This combination is important, as trades have an effect on the behavior of other traders and, thus, on the distribution of the LOB. Fourth, as illustrated with liquidity, the visualization is easily adjustable to include many variables. For example, it can display the informational content of the LOB, the behavior of HFTs in the market, changes in volatility, the moment(s) when price limits are hit or the cost of transactions over time. Moreover, variables can be visualized in various ways, as illustrated through trade volume and time. Fifth, two markets can be visualized at the same time to study their interconnectedness. For example, the LOB of wheat can be plotted along with that of corn, to study the effect of corn trades on the LOB of the wheat market. Moreover, it allows for cross-venue and cross-asset visualizations and all related markets can be visualized simultaneously: options, futures and spot markets for a specific product. Finally, the visualizations are created using ROOT, which has several advantages. It is multipurpose software that allows for more compact data storage and faster analyses. LOBs can be processed and analyzed relatively fast, which is beneficial for all financial stakeholders. Creating the visualizations is efficient and fast, and it allows users to interactively zoom, which will automatically update the plot. The technique is not limited to (agricultural) futures markets and can be used for any market with a LOB, such as equity, crypto and options markets.

The proposed methodology has several implications for financial market stakeholders. It offers an easy method to reconstruct the LOB and a useful tool for data exploration. As argued before, the visualization can be adjusted to study specific topics of interest. Academics can use the methodology to complement their research by adding visualizations with additional variables, such as the distribution of participants in the market, trade outliers or spoofing indicators. In addition, it is possible to visualize multiple markets in the same graph. For example, LOB volume differences at (similar) price levels can be used to study the varying LOB distributions in related markets. The synced LOBs facilitate future research in, among others, spillover effects, arbitrage trades and spread trading. This will help all parties involved to better understand markets. Regulators can use the methodology as a more advanced tool to monitor (related) markets and trace and study irregularities, such as market manipulation. It will help traders and controllers to monitor compliance. Furthermore, the methodology can help policymakers in regulating and designing the microstructure of markets. For example, the visualization can show the effect of larger and smaller price limits or adjustments on the tick size of the market. This is especially interesting when additional statistics are included in the visualization.

The proposed methodology carries several caveats. First, the visualization becomes hard to read when there are many messages. In this case, the program will skip the visualization of a predefined number of messages. For example, if there are 100 messages and the skip is set to ten messages, the $10^{th}$, $20^{th}$, $30^{th}$, etc. message will be plotted. However, due to the density of the messages, this is not visible to the eye. Second, adding multiple variables may be counterproductive, in that they may cluttering the visualization and thus make it harder to read and understand. Finally, academics and industry participants are used to seeing time being plotted on the x-axis, rather than in a separate graph. Therefore, it may take some adjustment and learning to get used to the new way of visualizing the LOB, where messages replace time on the x-axis.

From a methodological point of view, new metrics can be developed that can be applied to LOB visualization, for example, to measure the distribution of the LOB underlying the visualization. Future studies can explore whether certain distributions of the LOB are different from the "average" or benchmark LOB distribution. Furthermore, the use of particle anomaly detection may be helpful in tracing anomalies in LOB markets. Studies of irregularities and malpractices, such as spoofing, can apply this visualization technique to identify these anomalies. Future research on application aspects might focus on using this visualization method with different topics of interest, such as liquidity, volatility, market participant distribution, HFT behavior, large trades, market resilience, transaction costs, LOB information content, endogenous and exogenous shocks and spillover effects between markets.

## Data Availability Statement

The data that support the findings of this study are available from the Chicago Mercantile Exchange. Restrictions apply to the availability of these data, which were used under license for this study. The data are available from the authors with the permission of the Chicago Mercantile Exchange.

# References

Aidov, A., & Daigler, R. T. (2015). Depth Characteristics for the Electronic Futures Limit Order Book. *The Journal of Futures Markets*, *35*(6), 542–560. https://doi.org/https://doi.org/10.1002/fut.21706

Aldridge, I., & Krawciw, S. (2017). *Real-Time Risk: What Investors Should Know About FinTech, High-Frequency Trading, and Flash Crashes*. https://doi.org/10.1002/9781119319030

Antcheva, I., Ballintijn, M., Bellenot, B., Biskup, M., Brun, R., Buncic, N., … Tadel, M. (2009). ROOT — A C++ framework for petabyte data storage, statistical analysis and visualization. *Computer Physics Communications*, *180*(12), 2499–2512. https://doi.org/10.1016/j.cpc.2009.08.005

Arzandeh, M., & Frank, J. (2019). Price Discovery in Agricultural Futures Markets: Should We Look beyond the Best Bid-Ask Spread? *American Journal of Agricultural Economics*, *101*(5), 1482–1498. https://doi.org/10.1093/ajae/aaz001

Baruch, S., Panayides, M., & Venkataraman, K. (2017). Informed trading and price discovery before corporate events. *Journal of Financial Economics*, *125*(3), 561–588. https://doi.org/10.1016/j.jfineco.2017.05.008

Bayraktar, E., & Munk, A. (2018). Mini-Flash Crashes, Model Risk, and Optimal Execution. *Market Microstructure and Liquidity*, *04*(01n02), 1850010. https://doi.org/10.1142/S2382626618500107

Bhattacharya, U., Kuo, W.-Y., Lin, T.-C., & Zhao, J. (2018). Do Superstitious Traders Lose Money? *Management Science*, *64*(8), 3772–3791. https://doi.org/10.1287/mnsc.2016.2701

Biais, B., Bisière, C., & Spatt, C. (2010). Imperfect Competition in Financial Markets: An Empirical Study of Island and Nasdaq. *Management Science*, *56*(12), 2237–2250.

https://doi.org/10.1287/mnsc.1100.1243

Brogaard, J., & Garriott, C. (2019). High-Frequency Trading Competition. *Journal of Financial and Quantitative Analysis*, *54*(4), 1469–1497. https://doi.org/10.1017/S0022109018001175

Brogaard, J., Hendershott, T., & Riordan, R. (2014). High-Frequency Trading and Price Discovery. *Review of Financial Studies*, *27*(8), 2267–2306. https://doi.org/10.1093/rfs/hhu032

Brogaard, J., Hendershott, T., & Riordan, R. (2019). Price Discovery without Trading: Evidence from Limit Orders. *The Journal of Finance*, *74*(4), 1621–1658. https://doi.org/10.1111/jofi.12769

Brolley, M. (2020). Price Improvement and Execution Risk in Lit and Dark Markets. *Management Science*, *66*(2), 863–886. https://doi.org/10.1287/mnsc.2018.3204

Brun, R., & Rademakers, F. (1997). ROOT - an object oriented data analysis framework. *Nuclear Instruments and Methods in Physics Research Section A: Accelerators, Spectrometers, Detectors and Associated Equipment*, *389*(1–2), 81–86.

Buti, S., Rindi, B., & Werner, I. M. (2017). Dark pool trading strategies, market quality and welfare. *Journal of Financial Economics*, *124*(2), 244–265. https://doi.org/10.1016/j.jfineco.2016.02.002

Cao, C., Hansch, O., & Wang, X. (2009). The information content of an open limit-order book. *Journal of Futures Markets*, *29*(1), 16–41. https://doi.org/10.1002/fut.20334

Capponi, A., Menkveld, A. J., & Zhang, H. (2019). Large Orders in Small Markets: On Optimal Execution with Endogenous Liquidity Supply. *SSRN Electronic Journal*. https://doi.org/10.2139/ssrn.3326313

CERN. (2018a). About ROOT. Retrieved February 25, 2020, from https://root.cern.ch/about-root

CERN. (2018b). ROOT. Retrieved February 25, 2020, from https://root.cern

CERN. (2020a). Physics. Retrieved April 28, 2020, from https://home.cern/science/physics

CERN. (2020b). The Large Hadron Collider. Retrieved April 21, 2020, from https://home.cern/science/accelerators/large-hadron-collider

CERN. (2021a). RooFit. Retrieved February 8, 2021, from https://root.cern.ch/doc/master/group__Roofitmain.html

CERN. (2021b). RooStats. Retrieved February 8, 2021, from https://root.cern.ch/doc/v606/group__Roostats.html

CERN. (2021c). TMVA. Retrieved February 8, 2021, from https://root.cern.ch/tmva

CFTC-SEC. (2010). *Findings Regarding The Market Events of May 6, 2010*. Retrieved from https://www.sec.gov/news/studies/2010/marketevents-report.pdf

Chen, N., Kou, S., & Wang, C. (2018). A Partitioning Algorithm for Markov Decision Processes with Applications to Market Microstructure. *Management Science*, *64*(2), 784–803. https://doi.org/10.1287/mnsc.2016.2639

Chordia, T., Hu, J., Subrahmanyam, A., & Tong, Q. (2019). Order Flow Volatility and Equity Costs of Capital. *Management Science*, *65*(4), 1520–1551. https://doi.org/10.1287/mnsc.2017.2848

CME Group. (2020a). Develop to CME Globex. Retrieved March 12, 2020, from https://www.cmegroup.com/globex/develop-to-cme-globex.html

CME Group. (2020b). MDP 3.0 - Market by Price - Multiple Depth Book. Retrieved March 12, 2020, from https://www.cmegroup.com/confluence/display/EPICSANDBOX/MDP+3.0+-+Market+by+Price+-+Multiple+Depth+Book

Comerton-Forde, C., Malinova, K., & Park, A. (2018). Regulating dark trading: Order

flow segmentation and market quality. *Journal of Financial Economics*, *130*(2), 347–366. https://doi.org/10.1016/j.jfineco.2018.07.002

Du, S., & Zhu, H. (2017). What is the Optimal Trading Frequency in Financial Markets? *The Review of Economic Studies*, *84*(4), 1606–1651. https://doi.org/10.1093/restud/rdx006

Dugast, J. (2018). Unscheduled News and Market Dynamics. *The Journal of Finance*, *73*(6), 2537–2586. https://doi.org/10.1111/jofi.12717

Easley, D., & O'Hara, M. (1987). Price, trade size, and information in securities markets. *Journal of Financial Economics*, *19*(1), 69–90. https://doi.org/https://doi.org/10.1016/0304-405X(87)90029-8

Engle, R. F., & Russell, J. R. (1998). Autoregressive Conditional Duration: A New Model for Irregularly Spaced Transaction Data. *Econometrica*, *66*(5), 1127–1162. https://doi.org/10.2307/2999632

Erenburg, G., & Lasser, D. (2009). Electronic limit order book and order submission choice around macroeconomic news. *Review of Financial Economics*, *18*(4), 172–182. https://doi.org/https://doi.org/10.1016/j.rfe.2009.06.001

FIXtrading. (2020). Financial Information eXchange (FIX) Protocol. Retrieved February 25, 2020, from https://www.fixtrading.org/what-is-fix/

Gai, J., Choi, D. J., O'Neal, D., Ye, M., & Sinkovits, R. S. (2014). Fast construction of nanosecond level snapshots of financial markets. *Concurrency and Computation: Practice and Experience*, *26*(13), 2149–2156. https://doi.org/10.1002/cpe.3220

Gaillard, M. (2017). CERN Data Centre passes the 200-petabyte milestone. *CERN News*. Retrieved from https://home.cern/news/news/computing/cern-data-centre-passes-200-petabyte-milestone

Golub, A., Keane, J., & Poon, S.-H. (2012). High Frequency Trading and Mini Flash

Crashes. *SSRN Electronic Journal*. https://doi.org/10.2139/ssrn.2182097

Hachmeister, A. (2007). *Informed Traders as Liquidity Providers: Evidence from the German Equity Market*. https://doi.org/10.1007/978-3-8350-9577-9

Hasbrouck, J. (2017). Securities Trading: Principles and Procedures. *Manuscript, Version 12*, 1–195.

Hautsch, N., & Horvath, A. (2019). How effective are trading pauses? *Journal of Financial Economics*, *131*(2), 378–403. https://doi.org/10.1016/j.jfineco.2017.12.011

Hirsch, M., Cook, D., Lajbcygier, P., & Hyndman, R. (2019). Revealing High-Frequency Trading Provision of Liquidity with Visualization. *Proceedings of the 2nd International Conference on Software Engineering and Information Management*, 157–165.

Ito, T., & Yamada, M. (2018). Did the reform fix the London fix problem? *Journal of International Money and Finance*, *80*, 75–95. https://doi.org/10.1016/j.jimonfin.2017.10.004

Kandel, E., Rindi, B., & Bosetti, L. (2012). The effect of a closing call auction on market quality and trading strategies. *Journal of Financial Intermediation*, *21*(1), 23–49. https://doi.org/10.1016/j.jfi.2011.03.002

Kerr, J., Sadka, G., & Sadka, R. (2019). Illiquidity and Price Informativeness. *Management Science*, *66*(1), 334–351. https://doi.org/https://doi.org/10.1287/mnsc.2018.3154

Kirilenko, A., Kyle, A. S., Samadi, M., & Tuzun, T. (2017). The Flash Crash: High-Frequency Trading in an Electronic Market. *The Journal of Finance*, *72*(3), 967–998. https://doi.org/10.1111/jofi.12498

Kyle, A. S. (1985). Continuous Auctions and Insider Trading. *Econometrica: Journal of the Econometric Society*, *53*(6), 1315–1335. https://doi.org/10.2307/1913210

Li, S., Wang, X., & Ye, M. (2019). *Who Provides Liquidity, and When?* (No. 25972). https://doi.org/10.3386/w25972

LLVM. (2021). Clang: a C language family fronted for LLVM. Retrieved February 26, 2021, from https://clang.llvm.org/

Menkveld, A. J., & Yueshen, B. Z. (2019). The Flash Crash: A Cautionary Tale About Highly Fragmented Markets. *Management Science*, *65*(10), 4470–4488. https://doi.org/https://doi.org/10.1287/mnsc.2018.3040

Menkveld, A. J., & Zoican, M. A. (2017). Need for Speed? Exchange Latency and Liquidity. *The Review of Financial Studies*, *30*(4), 1188–1228. https://doi.org/10.1093/rfs/hhx006

Paddrik, M. E., Haynes, R., Todd, A. E., Scherer, W. T., & Beling, P. A. (2016). Visual Analysis to support regulators in electronic order book markets. *Environment Systems and Decisions*, *36*(2), 167–182. https://doi.org/https://doi.org/10.1007/s10669-016-9597-2

Peress, J., & Schmidt, D. (2020). Glued to the TV: Distracted noise traders and stock market liquidity. *The Journal of Finance*, *75*(2), 1083–1133. https://doi.org/https://doi.org/10.1111/jofi.12863

Rösch, C. G., & Kaserer, C. (2013). Market liquidity in the financial crisis: The role of liquidity commonality and flight-to-quality. *Journal of Banking & Finance*, *37*(7), 2284–2302. https://doi.org/10.1016/j.jbankfin.2013.01.009

Sinkovits, R., Feng, T., & Ye, M. (2014). Fast, Low-Memory Algorithm for Construction of Nanosecond Level Snapshots of Financial Markets. *Proceedings of the 2014 Annual Conference on Extreme Science and Engineering Discovery Environment - XSEDE '14*, 1–5. https://doi.org/10.1145/2616498.2616501

Tejedor, E., & Kothuri, P. (2018). CERN's Platform for Data Analysis with Spark.

Retrieved April 4, 2020, from https://www.slideshare.net/databricks/cerns-next-generation-data-analysis-platform-with-apache-spark-with-enric-tejedor

Trebbi, F., & Xiao, K. (2019). Regulation and Market Liquidity. *Management Science*, *65*(5), 1949–1968. https://doi.org/https://doi.org/10.1287/mnsc.2017.2876

Yao, C., & Ye, M. (2018). Why Trading Speed Matters: A Tale of Queue Rationing under Price Controls. *The Review of Financial Studies*, *31*(6), 2157–2183. https://doi.org/10.1093/rfs/hhy002

[dataset] CME Group; 2015; MDP 3.0 Market Data; https://www.cmegroup.com/market-data.html

# Appendix A

Particle physics studies the fundamental constituents of matter (CERN, 2020a). One of its most prominent institutions is the European Organization for Nuclear Research (CERN). At CERN, physicists conduct research using the Large Hadron Collider (LHC) (CERN, 2020b). Although the LHC and LOB are different in nature, they show similarities in terms of the data they generate and the analysis techniques they require. LHC data is highly granular, as particle properties are measured by hundreds of thousands of sensors, producing a detailed stream of values. LHC software reconstructs the particles' physical properties from these measured values. Almost everything at the LHC is obeying distributions: the initial collision (what collides and how?), the initial collision products (how did the particles react?), the properties of the particles flying through the detector (how did the initial particles transform?) and the sensors' measurement uncertainties. Due to these distributions, a single measurement is of very limited value: the wealth of the data is only accessible through statistical analyses. These analyses are performed on distributions of physical properties and their correlations, measuring the significance of how well models of the fundamental laws of physics (e.g. the Standard Model[9] of particle physics) describe the collision products and looking for deviations from these models.

Much of this also applies to financial data. The market microstructure and the strive of market participants to make a profit or manage their risks define boundary conditions for trading actions, producing causality similar to that described in the laws of physics. While each single action might appear to follow a random distribution, behavioral patterns emerge from statistical analysis of the ensemble of actions. Here, measurements of the LHC correspond to actions and indicators of the market.

---

[9] The Standard Model of particle physics is a theory that explains how the basic building blocks of the universe (fundamental particles) are related to fundamental forces. It is a well-tested theory, as it has successfully explained almost all experimental results and precisely predicted phenomena (CERN, 2020c).

A notable difference, however, is the time dependence of events. While particle collisions are stochastically independent from each other, i.e. any previous collision has no effect on a subsequent collision, financial events exhibit a high degree of (temporal) correlation. Nevertheless, some of the tools and many of the approaches are applicable to financial data, analysis and visualization, opening up new opportunities for gaining knowledge and understanding from financial data.

**Appendix B**

MDP data provides the market messages required to recreate the LOB with millisecond precision. Each file contains all contracts of the same futures contract, ordered by message number and time of arrival. A contract's data is spread across multiple files, with adjacent time windows within each file. Each file contains three types of messages. Two types of messages are used for initialization and metadata (i.e. the "definition message" and the "security status message" respectively), and one is used for the incremental LOB updates. The definition and security status messages only exist once per contract per file. They contain information such as ID, name, expiration date, order book depth and tick size. The message that is used for incremental LOB updates consists of many different types of submessages. The main six submessages are messages to update the order book, i.e. insert a new bid or ask level, change existing bid or ask levels and delete existing bid or ask levels. Furthermore, there is one submessage that indicates when a trade took place and ten more submessages containing statistics such as opening price and settlement price (CME Group, 2020b).

To recreate the LOB, all messages are read and orders are processed in an iterative process. First, an empty order book of depth ten is initiated. Next, ten insertion submessages are initiated for the bid side and another ten for the ask side. Together, these twenty messages generate the first state of the LOB. Submessages that record changes in volume (at a certain level) replace the value of the quantity at the relevant price level entirely, meaning that these submessages contain the new actual quantity, rather than the volume that needs to be added or subtracted from the current quantity. If the volume at a certain level reaches zero, this level is deleted from the LOB with the delete-level submessage. If new volume is added to a previously non-existent level (zero volume), an insert-level submessage is sent (CME Group, 2020b).